\newcommand{\et}{\hat{\bf e}_\theta}

\newcommand{\ignore}[1]{}
\documentstyle[prb,aps,multicol,psfig]{revtex}
\begin{document}
\draft
\title{Resistive axisymmetric equilibria with arbitrary flow}

\author{M.\ P.\ Bora\footnote{Permanently at \it Physics Department, 
Gauhati University,
Guwahati, Assam 781~014, India.\\
Electronic mail~:~{\tt mbora@gw1.dot.net.in}}}
\address{Institute for Plasma Research, Bhatt, Gandhinagar 382~428, India.}

\maketitle

\begin{abstract}
An analysis of axisymmetric equilibria with arbitrary incompressible flow
and finite 
resistivity is presented.  It is shown that with large aspect ratio approximation or 
vanishing poloidal current, a uniform conductivity profile is consistent with 
equilibrium flows. Also a comment made on  coexistence of both toroidal and 
poloidal flows in an axisymmetric field-reversed configuration.

\end{abstract}

\pacs{PACS numbers~:~52.30.Bt, 52.30.Jb, 52.55.-s}


\begin{multicols}{2}

Calculating the equilibrium is one of the fundamental problems of magnetically 
confined plasmas. Most studies are directed toward finding an ideal (i.e.\ 
infinitely conducting) magnetohydrodynamic (MHD) equilibria in an axisymmetric 
plasma. The earliest calculations are those of Grad \cite{grad1} and Shafranov 
\cite{shafranov1}, leading to the famous Grad-Shafranov equation 
\cite{grad1,shafranov1,lust1}. The ideal and static Grad-Shafranov equation is 
an elliptic differential equation in the magnetic flux function $\psi$ with two 
arbitrary surface quantities as the pressure $p(\psi)$ and the poloidal current 
$I(\psi)$.

Consequently, there have been attempts to include various 
effects e.g.\ mass flow into the equilibrium equations. An equivalent of 
Grad-Shafranov equation in an 
axisymmetric ideal plasma with arbitrary flow has been given by Hameiri 
\cite{hameiri1}. In some recent works, Steinhauer \cite{stein1} deals with a 
generalization of Grad-Shafranov equilibria in a multi-fluid with flow and 
Throumoulopoulos and Tasso \cite{throm4} consider a helically symmetric 
equilibria with flow. The situation with flow becomes more realistic when one 
realizes the existence of equilibrium flows both in toroidal and poloidal 
directions in tokamaks following momentum deposition through heating by neutral 
beam injection \cite{suckewer1,brau1,scott1}. With equilibrium flows, the 
resultant governing differential equation does not remain always elliptic 
\cite{throm1,semenzato1}. 
The investigation of a general MHD equilibrium becomes much more 
complicated when one tries to include the effects of other important factors, 
say of viscous stress tensor. 
Recently Ren et al.\ \cite{ren1} have studied the deformation of magnetic island 
by including the effect of sheared flow and viscosity into an ideal 
two-dimensional 
MHD equilibrium configuration. However, there is an element of inconsistency, 
whether 
an ideal equilibrium is realistic \cite{mont1}. Heuristically, one ignores the 
resistivity in the Ohm's law while calculating the equilibrium, but then a
resistive stability analysis based on a stationary equilibrium remains 
questionable as long as the field diffusion is not taken into 
account \cite{dobrott1}.
Montgomery et al.\cite{mont1} have investigated the problem on non-ideal static 
axisymmetric equilibrium. There have also been attempts to calculate resistive 
axisymmetric equilibrium with only toroidal flow \cite{throm1}. It has been 
further argued that tokamak equilibrium flow is either purely toroidal 
\cite{suckewer1,scott1} or the poloidal component is small \cite{brau1} and 
quickly damped by magnetic pumping \cite{hassam1}. So there is a natural 
tendency to exclude the poloidal flow while calculating an equilibrium. But when 
one considers finite conductivity with purely equilibrium toroidal flow, the 
conductivity (hence the resistivity) becomes a function of space. In general, 
the resistivity is not a flux function \cite{throm1,mont1} irrespective of 
equilibrium flow. In this report, we ask the very pertinent question, whether 
the situation changes in presence of poloidal flow. As we show that a 
uniform resistivity profile is consistent in presence of poloidal flow, 
whereas it has been shown that a scalar pressure equilibrium can not have 
uniform 
resistivity \cite{bates1}. Further we show that in a field-reversed (FRC) 
axisymmetric 
configuration with no toroidal magnetic field, both toroidal and poloidal 
equilibrium flows can coexist with finite resistivity, which is not found to be 
the case with ideal equilibrium \cite{clemente1}.

We consider the equilibrium resistive MHD equations with plasma flow. The
equations are
\begin{eqnarray}
\nabla\cdot(\rho{\bf v}) &=& 0,
\label{continuity}\\
\nabla\cdot{\bf B} = 0, \nabla\times{\bf E} &=& 0, \nabla\times{\bf B} = 0,
\label{induction}\\
{\bf E}+{\bf v}\times{\bf B} &=& {\bf j}/\sigma,
\label{ohm}\\
\rho({\bf v}\cdot\nabla){\bf v} &=& {\bf j}\times{\bf B}-\nabla p,
\label{momentum}
\end{eqnarray}
where the symbols have their usual meanings.
We use a right handed cylindrical system $(r,\theta,z)$ with $z$ as the axis
of symmetry, $\theta$ as the toroidal angle, and $r$ along the major radius
of an axisymmetric device. We assume the plasma flow to be arbitrary (toroidal 
and poloidal) and 
axisymmetry is assumed i.e.\
$\partial/\partial\theta=0$. The plasma resistivity $\eta=\sigma^{-1}$ is 
assumed to
be an unspecified function of $r$ and $z$. We have further assumed here that
the equilibrium is maintained in a steady-state through resistive diffusion.
The magnetic induction equation 
allows us to write the magnetic field as
\begin{equation}
{\bf B}={1\over r}\nabla\psi\times\et+{I\over r}\et,
\label{bdefn}
\end{equation}
where $\psi$ is the magnetic flux function which is the azimuthal component
of the vector potential ${\bf A}$ and $I$ is the current function.
Similarly, following the continuity equation,
Eq.(\ref{continuity}), we can express the plasma equilibrium velocity as
\begin{equation}
{\bf v}={1\over\rho r}\nabla\varphi\times\et+\omega r\et,
\label{vdefn}
\end{equation}
where $\varphi$ is the velocity stream function and $\omega=v_\theta/r$ is the
toroidal angular velocity. We also assume that the flow is incompressible
i.e.\ $\nabla\cdot{\bf v}=0$.

Because the flow is now in both toroidal and poloidal direction, the poloidal
component of current, ${\bf j}_p$ need not vanish. In general the current can
be expressed as
\begin{equation}
{\bf j}=-{1\over r}\Delta^\ast\psi\et+{1\over r}\nabla I\times\et,
\label{jdefn}
\end{equation}
where $\Delta^\ast$ is the elliptic operator defined by
%
$\Delta^\ast\psi=r^2\nabla\cdot\left({1\over r^2}\nabla\psi\right).$
%
Taking curl of the Ohm's law Eq.(\ref{ohm}), we have
\begin{equation}
\nabla\times({\bf v}\times{\bf B})=\nabla\times({\bf j}/\sigma)
\label{curl_vb}
\end{equation}
with
\begin{equation}
{\bf v}\times{\bf B}={1\over\rho r^2}\nabla\varphi\times\nabla\psi-
{I\over\rho r^2}\nabla\varphi+\omega\nabla\psi.
\label{vcrossb}
\end{equation}
The $\et$ component of Eq.(\ref{curl_vb}) can be now written as
\begin{eqnarray}
\et\cdot\nabla\omega\times\nabla\psi-\et\cdot\nabla\left({I\over\rho r^2}
\right)\times\nabla\varphi
\nonumber\\
={1\over\sigma r}\left({1\over\sigma}\nabla\sigma\cdot\nabla I
+{2\over r}{\partial I\over\partial r}-\nabla^2I\right).
\end{eqnarray}

We now invoke the large aspect ratio expansion and assume that toroidal magnetic
field, to the first approximation can be written $B_\theta\simeq B_0r_0/r$. 
Here,
$B_0$ is the value of the toroidal magnetic field at center of the cylindrical
cross section of the torus at distance $r_0$ from the axis of symmetry. To this
effect we have the current function $I\approx B_0r_0=\rm const$. Under these
assumption, the above equation can be written as
\begin{eqnarray}
\et\cdot\nabla\omega\times\nabla\psi &=& 
-{I\over\rho^2r^2}\et\cdot\nabla\rho\times
\nabla\varphi
\nonumber\\
&& -\, {2I\over\rho r^3}\et\cdot\nabla r\times\nabla\varphi.
\end{eqnarray}
The first term on the right hand side of the above equation vanishes
by virtue of the incompressibility condition and the continuity 
equation. We neglect the second term because of its ${1/r^3}$ dependence
and find that the toroidal angular velocity $v_\theta/r=\omega(\psi)$
becomes a surface quantity. This also further means that ${\bf j}=
j_\theta\et$ with ${\bf j}_p=0$. We, however, note that the condition
$\omega\equiv\omega(\psi)$ is identically satisfied in a field-reversed
configuration (FRC) where $I=0$ and large aspect ratio approximation
is not required. It can be noted here that without any approximation, $\omega$ 
becomes a flux function when one considers ideal equilibrium \cite{hameiri1} or 
resistive equilibrium with only toroidal flow \cite{throm1}.

Now, we consider the momentum equation Eq.(\ref{momentum}) and its
$\et$ component. With the above approximations, we can write 
Eq.(\ref{momentum}) as
\begin{eqnarray}
j_\theta\et\times{\bf B}_p &=& \nabla P+
\rho\nabla\left[{1\over2\rho^2r^2}(\nabla\varphi)^2\right]
\nonumber\\
&& -\, 
\nabla\cdot\left({1\over\rho r^2}\nabla\varphi\right)\nabla\varphi
-\omega'(\nabla\varphi\times\nabla\psi)
\nonumber\\
&& -\,\,{2\omega\over
 r}{\partial\varphi\over\partial z}\et-\omega^2r\nabla r
\label{vdelv}
\end{eqnarray}
In the above equation ${\bf B}_p$ is the poloidal component of the magnetic
field and $(')$ denotes derivative with respect to $\psi$. Taking the $\et$
component of the above equation, we have,
\begin{equation}
\et\cdot\nabla\varphi\times\nabla(\omega r^2)=0,
\label{phi_om}
\end{equation}
which means that $\varphi\equiv\varphi(\omega r^2)$. We take the simplest
situation of $\varphi\propto\omega r^2$ which yields another surface
quantity, $\varphi/r^2=\zeta(\psi)$. However, it is important to note that,
$\zeta(\psi)$ is not an arbitrary function in the sense that it is proportional
to the toroidal velocity $\omega(\psi)$ i.e.\ the toroidal and poloidal flows
are no longer independent. Physically, one can understand this by noting that
finite resistivity allows plasma motion across the the flux surfaces.
Equivalently, toroidal flow, in a resistive axisymmetric plasma, is always
associated with poloidal flow.

Because of
equilibrium flow, however, plasma pressure $p$ is no longer a flux
function now. Taking the ${\bf B}_p$ component of the momentum equation
(Eq.(\ref{vdelv})), we have
\begin{eqnarray}
{\bf B}_p\cdot\left[{\nabla p\over\rho}+\nabla\left\{{1\over2\rho^2r^2}
(\nabla\varphi)^2-{1\over2}\omega^2r^2\right\}\right]
\nonumber\\
={1\over\rho}\nabla
\cdot\left({1\over\rho r^2}\nabla(\zeta r^2)\right){\bf B}_p\cdot
\nabla(\zeta r^2).
\label{bpeqn}
\end{eqnarray}

Depending upon the equation of state, now, several options are possible.
However, we note that density, in general, is not a flux function in
presence of arbitrary plasma flow. this can be easily seen from the
equation of continuity Eq.(\ref{continuity}), 
after applying the 
incompressibility condition,
\begin{equation}
\et\cdot\nabla\varphi\times\nabla\rho=0.
\label{phi_rho}
\end{equation}
It can be seen from the above expression that $\rho$ is not a surface
quantity. We note here that axisymmetric 
equilibria with incompressible equilibrium flows are generally 
associated with constant density magnetic 
surfaces \cite{avinash1,throm2,throm3}. One is also free to choose density as a 
flux function in case of resistive axisymmetric equilibrium with only 
incompressible toroidal 
flow \cite{throm1}.

Taking the $\et$ component of Ohm's law Eq.(\ref{ohm}) along with
Eq.(\ref{vcrossb}), we have an expression for plasma conductivity,
\begin{equation}
\sigma\left(E_0r_0+{2\over\rho}r\zeta B_r\right)+\Delta^\ast\psi=0,
\label{conductivity}
\end{equation}
where $E_0$ is the longitudinal externally applied electric field
at major radius $r=r_0$.
We immediately see from the above expression that conductivity, in general, is
a space dependent quantity.

In what follows, we shall consider two cases with ({\it i\/})~uniform and
constant density
and ({\it ii\/})~a nonuniform density. In the second case, we consider
isentropic magnetic surfaces.
We now assume that plasma density is uniform and constant i.e.\ 
$\rho=\rm const.$ and normalize our equations to $\rho=1$. We can
now write Eq.(\ref{bpeqn}) as,
\begin{eqnarray}
{\bf B}_p\cdot\nabla\left[p+{1\over2r^2}(\nabla\varphi)^2-
{1\over2}\omega^2r^2\right]
\nonumber\\
={1\over r^2}
\Delta^\ast(\zeta r^2){\bf B}_p\cdot\nabla(\zeta r^2).
\end{eqnarray}
Integration of the above equation yields the equivalent Bernoulli's
equation,
\begin{eqnarray}
p+{(\nabla\varphi)^2\over2r^2}-{\omega^2r^2\over2} &=&
\int\!\!{dl\over B_p}\, {1\over r^2}\Delta^\ast(\zeta r^2){\bf B}_p
\cdot\nabla(\zeta r^2)
\nonumber\\
&& +\, \chi(\psi),
\label{bernoulli1}
\end{eqnarray}
where the integration is along a magnetic field line and $\chi(\psi)$
is an arbitrary surface quantity.
The solubility condition further requires that
\begin{equation}
\oint{dl\over B_p}\, {1\over r^2}\Delta^\ast(\zeta r^2){\bf B}_p
\cdot\nabla(\zeta r^2)=0.
\end{equation}
We further assume that part of the pressure gradient that varies within
a magnetic flux tube has no $\nabla\psi$ component \cite{clemente1} i.e.
\begin{equation}
\nabla\psi\cdot\nabla\oint{dl\over B_p}\, 
{1\over r^2}\Delta^\ast(\zeta r^2){\bf B}_p
\cdot\nabla(\zeta r^2)=0.
\label{delpsi}
\end{equation}
Together with Eq.(\ref{bernoulli1}) and the above assumption, the 
$\nabla\psi$ component of the momentum equation yields the equivalent
Grad-Shafranov equation,
\begin{equation}
\Delta^\ast\psi+r^2(\chi'+\omega'\omega^2r)+{
\Delta^\ast(\zeta r^2)\nabla\psi\cdot\nabla(\zeta r^2)\over|\nabla\psi|^2}=0,
\label{grad1}
\end{equation}
with two arbitrary flux functions $\chi(\psi)$ and $\omega(\psi)$.
The primes refer derivative with respect to $\psi$.

We now consider the second case where we consider a nonuniform density.
With incompressible flow, magnetic surfaces with constant entropy is
quite a reasonable approximation in ideal MHD \cite{hameiri1,morozov1}.
However, 
considering 
long resistive diffusion time, the right hand side of Eq.(\ref{ohm})
can be neglected and we can continue to proceed with isentropic magnetic
surfaces \cite{throm1}. The equation of state can now be written as,
%
$p=S\rho^\gamma$,
%
where, $S(\psi)$ is the entropy which is a flux function and $\gamma$
is the ratio of specific heats. We now write ${\bf B}_p\cdot\nabla p/\rho$
as ${\bf B}_p\cdot\nabla[\gamma S\rho^{\gamma-1}/(\gamma-1)]$, so that
equivalent Bernoulli's equation can be written as,
\begin{eqnarray}
\Theta(\psi)+
\int{dl\over B_p}\, {1\over\rho}\nabla\cdot\left({1\over\rho r^2}
\nabla(\zeta r^2)\right){\bf B}_p\cdot\nabla(\zeta r^2)\nonumber\\
={\gamma\over\gamma-1}S\rho^{\gamma-1}+{1\over2\rho^2r^2}(\nabla\varphi)^2
-{1\over2}\omega^2r^2,
\label{bernoulli2}
\end{eqnarray}
where $\Theta(\psi)$ is arbitrary. As we have assumed previously, it 
requires a solubility condition and the equivalent to the assumption
(\ref{delpsi}). We can then continue to write the equivalent Grad-Shafranov
equation by taking the $\nabla\psi$ component of the momentum equation and
applying the Bernoulli's law Eq.(\ref{bernoulli2}),
\begin{eqnarray}
\Delta^\ast\psi+r^2\left(\Theta'+\omega'\omega^2r-
S'{\rho^{\gamma-1}\over\gamma-1}\right)
\nonumber\\
={r^2\over\rho|\nabla\psi|^2}\nabla\cdot
\left[{1\over\rho r^2}\nabla(\zeta r^2)\right]\nabla\psi\cdot\nabla
(\zeta r^2).
\label{grad2}
\end{eqnarray}
In the above equation we have four arbitrary surface quantities i.e.\ 
$\Theta(\psi)$,
$\omega(\psi)$, and $S(\psi)$ and the primes denote
derivative with respect to $\psi$.

We have derived the differential equations, equivalent to the Grad-Shafranov 
equation, for resistive axisymmetric plasma with arbitrary equilibrium flows. 
These equilibrium equations Eqs.(\ref{grad1}, \ref{grad2}) have to be 
solved subject to conductivity constraint Eq.(\ref{conductivity}).
Further, in a field-reversed configuration (FRC) with no toroidal magnetic 
field, it can 
be seen from Eq.(\ref{phi_om}) that both poloidal and toroidal flow can coexist.

We now show that a uniform conductivity profile is consistent with resistive 
axisymmetric equilibria with arbitrary flow. A simple examination of 
Eq.(\ref{momentum}), though reveals that uniform conductivity may be possible
with scalar pressure equilibrium in presence of flow, it however provides no 
easier way of proving it. We note that the usual procedure for solving
Eqs.(\ref{grad1}) and (\ref{grad2}) requires specifying {\it a priori\/}
dependence of the respective arbitrary functions on $\psi$. However, in the
presence of finite resistivity, the resistivity constraint 
Eq.(\ref{conductivity}) can be used to solve for $\psi$, which is uniquely
determined if the right hand side of Eq.(\ref{conductivity}) is specified
\cite{bates1}. It should be noted here with caution whether the resultant
solution for $\psi$ corresponds to realistic profiles for other physical
quantities such as pressure, density, velocity etc. However, our sole aim,
here, is to demonstrate the existence of a solution consistent with uniform
resistivity in presence of flows.

From Eq.(\ref{phi_rho}) we know that $\rho\equiv\rho(\varphi)$, and assume that 
$\rho\propto\varphi$. We now assume that conductivity is uniform in space so 
that
the resulting Eq.(\ref{conductivity}) can be written as,
\begin{equation}
\Delta^\ast\psi+{\alpha\over r^2}{\partial\psi\over\partial z}=\beta,
\label{conductivity1}
\end{equation}
\end{multicols}

\begin{figure}
\centerline{\psfig{file=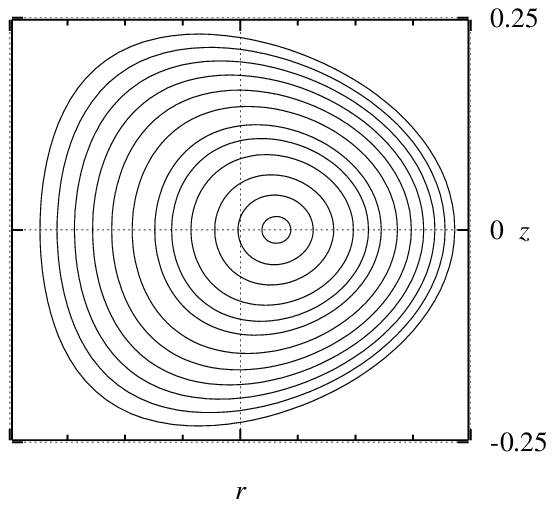}\hskip1.5in
\psfig{file=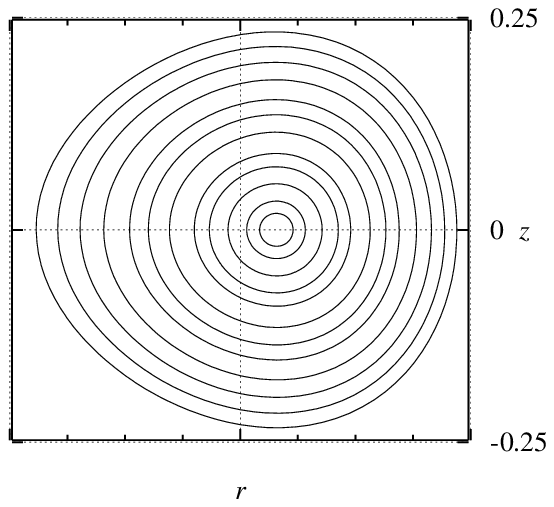}}
\caption{Constant flux ($\psi$) contours for conducting circular 
boundary with ({\it a\/})~$\sigma=\rm const.$ and ({\it b\/})~$\sigma\propto 
r^2$. The major radius is $r_0=1.25$.}
\end{figure}

\begin{multicols}{2}
\noindent
where $\alpha$ and $\beta$ are arbitrary constants. It is worthwhile mentioning
at this point that Eq.(\ref{conductivity1}) can not be used in case of
very small resistivity. In the limit of vanishing resistivity (large $\beta$ in
the above equation), the solution of Eq.(\ref{conductivity1}) contains short
scale spatial dependence (boundary layers), not present in case of ideal
equilibrium and may lead to unphysical results. 

Note that the above equation 
is a elliptic equation and can be treated as boundary value problem. Following 
Zheng et al.\ \cite{zheng1}, we assume a solution of the form
\begin{equation}
\psi_h(r,z)=\sum_{n=0,1,2,\ldots}f_n(r)z^n
\label{psi_hom}
\end{equation}
%
%
for the homogeneous part of Eq.(\ref{conductivity1}). We however retain the odd 
terms in the summation to take care of the asymmetric-term in 
Eq.(\ref{conductivity1}). For simplicity we assume that $f_n(r)=0$ for $n\geq3$, 
which, however, can be extended up to any number of terms if required, about 
which we shall make a comment later. Substituting Eq.(\ref{psi_hom}) in the 
equivalent homogeneous equation for Eq.(\ref{conductivity1}) we can solve for 
the 
functions $f_{0,1,2}(r)$. The homogeneous solution of Eq.(\ref{conductivity1}) 
is 
then given by,
\begin{eqnarray}
\psi_h(r,z) &=& a_1r^2\{4r^2+16z^2-\alpha^2[4(\ln r)^2-4\ln r+2
\nonumber
\\
&& -\, r^2]+8\alpha z(r^2-2\ln r)\}+a_2[2r^2(2\ln r
\nonumber
\\
&& -\, 1)-\alpha^2\ln r(\ln r+1)+4\alpha
z\ln r+4z^2]
\nonumber
\\
&& +\, a_3r^2[\alpha(2\ln r-1)+4z]+a_4r^2,
\label{psi_soln}
\end{eqnarray}
where $a_i$s are arbitrary constants to be determined from the boundary 
conditions. A particular solution of Eq.(\ref{conductivity1}) is 
$\psi_p=\beta r^2(2\ln r-1)/4$. So the complete solution of 
Eq.(\ref{conductivity1}) is
\begin{equation}
\psi=\psi_h+\psi_p,
\label{psi_soln1}
\end{equation}
which can be verified by direct substitution. For a conducting circular boundary 
of a toroidal axisymmetric device, the constant flux ($\psi$) contours are shown 
in Fig.1 ({\it a\/}) which shows a scaler pressure equilibrium. The solution for 
$\sigma\propto r^2$ is shown in Fig.1 ({\it b\/}). Note that $\sigma\propto r^2$ 
is the only possible solution for resistive axisymmetric equilibrium without 
flow \cite{bates1}. 

In principle the expansion in Eq.(\ref{psi_soln}) should be 
retained with a large number of terms which will result a equally large number 
of arbitrary constants for the solution in $\psi$. These constants can then be 
used to shape any arbitrarily shaped plasma boundary.

In passing, we would like to note that resistive field diffusion
($\partial B/\partial t\neq0$) is intrinsically involved with non-stationary 
equilibria ($v\neq0$). However, a series of ideal quasi-stationary equilibrium
states can
be built up with $\partial B/\partial t=0$ in which, the effect of finite
resistivity is only to slowly evolve the equilibrium in a diffusive time scale
\cite{grad3}.

It is a pleasure to thank A. Sen for the kind hospitality at IPR where part of
this work has been completed.

\end{multicols}

%
%
%
%

\begin{references}

\bibitem{grad1} H. Grad and H. Rubin, in {\it Proceedings of the Second
United Nations Conference on the Peaceful Uses of Atomic Energy, Geneva, 1958},
edited by United Nations (United Nations Publications, Geneva, 1958),
Vol.~{\bf 31}, pp.~190.

\bibitem{shafranov1} V. D. Shafranov, Sov.\ Phys.\ JETP {\bf 6}, 545 (1958).

\bibitem{lust1} R. L\"ust and A. Schl\"uter, Z.\ Naturforsch {\bf 12a}, 850
(1957).

\bibitem{hameiri1} E. Hameiri, Phys.\ Fluids {\bf 26}, 230 (1983).

\bibitem{stein1} L. C. Steinhauer, Phys.\ Plasmas {\bf 6}, 2734 (1999).

\bibitem{throm4} G. N. Throumoulopoulos and H. Tasso, ``Ideal
magnetohydrodynamic equilibria with helical symmetry and incompressible
flows'', e-print physics/9907004 (to be published in J.\ Plasma Phys.).
\bibitem{suckewer1} S. Suckewer, H. P. Eubank, R. J. Goldston {\it et al}.,
Phys.\ Rev.\ Lett.\ {\bf 43}, 207 (1979).

\bibitem{brau1} K. Brau, M. Bitter, R. J. Goldston {\it et al}., Nucl.\ Fusion
{\bf 23}, 1643 (1983).
\bibitem{scott1} S. D. Scott, M. Bitter, H. Hsuan {\it et al}., in {\it
Proceedings of the 14st European Conference on Controlled Fusion and Plasma
Physics, Madrid, 1987} (European Physical Society, Geneva 1987), Vol.~{\bf
11D}, pp.~65.
\bibitem{throm1} G. N. Throumoulopoulos, J.\ Plasma Phys.\ {\bf 59}, 303 (1998).

\bibitem{semenzato1} S. Semenzato, R. Gruber, and H. P. Zehrfeld, Comp.\ Phys.\
Rep.\ {\bf 1}, 389 (1984).

\bibitem{ren1} C. Ren, M. S. Chu, and J. D. Callen, Phys. Plasmas {\bf 6}, 1203
(1999).

\bibitem{mont1} D. Montgomery, J. W. Bates, and H. R. Lewis, J.\ Plasma Phys.\
{\bf }, (1997).

\bibitem{dobrott1} D. Dobrott, S. C. Prager, and J. B. Taylor, Phys.\ of
Fluids {\bf 20}, 1850 (1977).

\bibitem{hassam1} A. B. Hassam and R. M. Kulsrud, Phys.\ Fluids {\bf 21}, 2271
(1987).

\bibitem{bates1} J. W. Bates and H. R. Lewis, Phys.\ Plasmas {\bf 3}, 2395
(1996).

\bibitem{clemente1} R. A. Clemente and R. L. Viana, Plasma Phys.\ Control.\
Fusion {\bf 41}, 567 (1999).

\bibitem{avinash1} K. Avinash, S. N. Bhattacharyya, and B. J. Green, Plasma
Phys.\ Control.\ Fusion {\bf 34}, 465 (1992).

\bibitem{throm2} G. N. Throumoulopoulos and G. Pantis, Plasma Phys.\ Control.\
Fusion {\bf 38}, 1817 (1996).

\bibitem{throm3} G. N. Throumoulopoulos and H. Tasso, Phys.\ Plasmas {\bf 4},
1492 (1997).

\bibitem{morozov1} A. I. Morozov and L. S. Solov\'ev, in {\it Reviews of Plasma
Physics}, edited by M. A. Leontovich (Consultants Bureau, New York, 1980),
Vol.~{\bf 8}, pp.~1.

\bibitem{zheng1} S. B. Zheng, A. J. Wootton, and R. Solano, Phys.\ Plasmas {\bf 
3}, 1176 (1996).

\bibitem{grad3} H. Grad and J. Hogan, Phys.\ Rev.\ Lett.\ {\bf 24}, 1337 (1979).

\end{references}
\end{document}